# Initialization of a spin qubit in a site-controlled nanowire quantum dot

Konstantinos G. Lagoudakis[1*], Peter L. McMahon[1*], Kevin A. Fischer[1], Shruti Puri[2], Kai Müller[1], Dan Dalacu[3], Philip J. Poole[3], Michael E. Reimer[4], Val Zwiller[5], Yoshihisa Yamamoto[1,6], Jelena Vučković[1]

1 E.L. Ginzton Laboratory, Stanford University, Stanford, California 94305, USA

2 Département de Physique, Université de Sherbrooke, Sherbrooke, J1K 2R1 Québec, Canada

3 National Research Council of Canada, Ottawa, K1A 0R6, Canada

4 Institute for Quantum Computing and Department of Electrical and Computer Engineering, University of Waterloo, Waterloo, Ontario, N2L 3G1 Canada

5 Quantum Nano Photonics, KTH Royal Institute of Technology, SE-100 44 Stockholm, Sweden

6 National Institute of Informatics, Hitotsubashi 2-1-2, Chiyoda-ku, Tokyo 101-8403, Japan

* These authors contributed equally to this work.

**Abstract.** A fault-tolerant quantum repeater or quantum computer using solid-state spin-based quantum bits will likely require a physical implementation with many spins arranged in a grid. Self-assembled quantum dots (QDs) have been established as attractive candidates for building spin-based quantum information processing devices, but such QDs are randomly positioned, which makes them unsuitable for constructing large-scale processors. Recent efforts have shown that quantum dots embedded in nanowires can be deterministically positioned in regular arrays, can store single charges, and have excellent optical properties, but so far there have been no demonstrations of spin qubit operations using nanowire quantum dots. Here we demonstrate optical pumping of individual spins trapped in site-controlled nanowire quantum dots, resulting in high-fidelity spin-qubit initialization. This represents the next step towards establishing spins in nanowire quantum dots as quantum memories suitable for use in a large-scale, fault-tolerant quantum computer or repeater based on all-optical control of the spin qubits.

**Introduction:** The development of site-controlled QDs, and demonstration of their suitability for hosting spin-based qubits, is a key objective in the roadmap towards a scalable quantum information processing system implemented with quantum dots[1]-[3]. There has been considerable recent effort in exploring different techniques for fabricating site-controlled quantum dots, including lithographic patterning of growth substrates[4],[5], stress-induced positioning within micropillars[6], and the growth of quantum dots within seeded nanowires[7]. Quantum dots within nanowires have been shown to have both high photon extraction efficiencies[8],[9], and good single-photon source characteristics[7]-[9]. Furthermore, magneto-photoluminescence spectroscopy studies[10],[11] of InAsP QDs in InP nanowires have shown that QDs in nanowires may be a promising platform for hosting spin qubits, but to our knowledge, thus far there have been no demonstrations of the fundamental spin manipulation operations[12]-[20] on spins trapped in nanowire-hosted QDs, nor in any other site-controlled QD devices. Here, we demonstrate all-optical initialization of a spin qubit embedded in a deterministically positioned InP nanowire QD.

**Methods:** We studied a sample with InAsP QDs embedded in InP nanowires that was grown using vapor-liquid-solid epitaxy on a (111)B InP substrate; the growth details can be found in Ref. [7]. The substrate was covered with a $SiO_2$ mask containing a grid of apertures, which were produced using e-beam nanopatterning. The growth of each nanowire was seeded by placing a gold nanoparticle in the center of each aperture[7],[21],[22] and consisted of a two-step process that involves growth of a core nanowire containing the InAsP QD followed by growth of a shell, which results in the needle-like shape of the nanowires.

Photoluminescence spectra were measured using a custom double-grating spectrometer setup with ~10 μeV resolution, which is necessary in order to spectrally select just a single emission line from the QD, and measure its signal on a single-photon counter, while rejecting any laser light that may be near- or on-resonance with one of the other QD lines.

For all the experiments that required application of a magnetic field, we used the Voigt geometry, since this is the configuration that is used for spin control with optical pulses[15],[16],[18], and for generation of spin-photon entanglement[23]-[26].

**Results:** In this article we present results from a typical quantum-dot--nanowire device in our sample; the QD exhibited emission under CW above-band (780 nm) excitation that was both bright and spectrally narrow. Figure 1a shows the polarization-resolved spectra from the quantum dot. At saturation the quantum dot exhibited a linewidth of ~60 μeV full-width-at-half-maximum, and an energy difference of δE=1.6 μeV between the two orthogonal linear polarization components (measured between the fitted peak centers; see Figure 1a). The emission intensity of both H- and V-polarized lines have a linear dependence on the above-band laser power, until saturation is reached at approximately 550 nW (Figure 1b). The linear power dependence is consistent with these lines corresponding to single exciton emission, as opposed to biexciton emission (which would exhibit a quadratic power dependence). To determine that the quantum dot was charged, we used magneto-photoluminescence spectroscopy in the Voigt configuration (magnetic field perpendicular to the nanowire growth direction): Figure 1c shows the photoluminescence signal from the quantum dot as a function of its emission energy and polarization. The photoluminescence spectra clearly show a four-fold splitting, which is consistent with emission from a charged dot[27]. We note that in this figure and the remainder of the paper, we define H as the linear polarization that is θ=50º relative to the magnetic field and V as the linear polarization that is at θ=140º relative to the magnetic field, as shown in the illustration in Figure 1d.

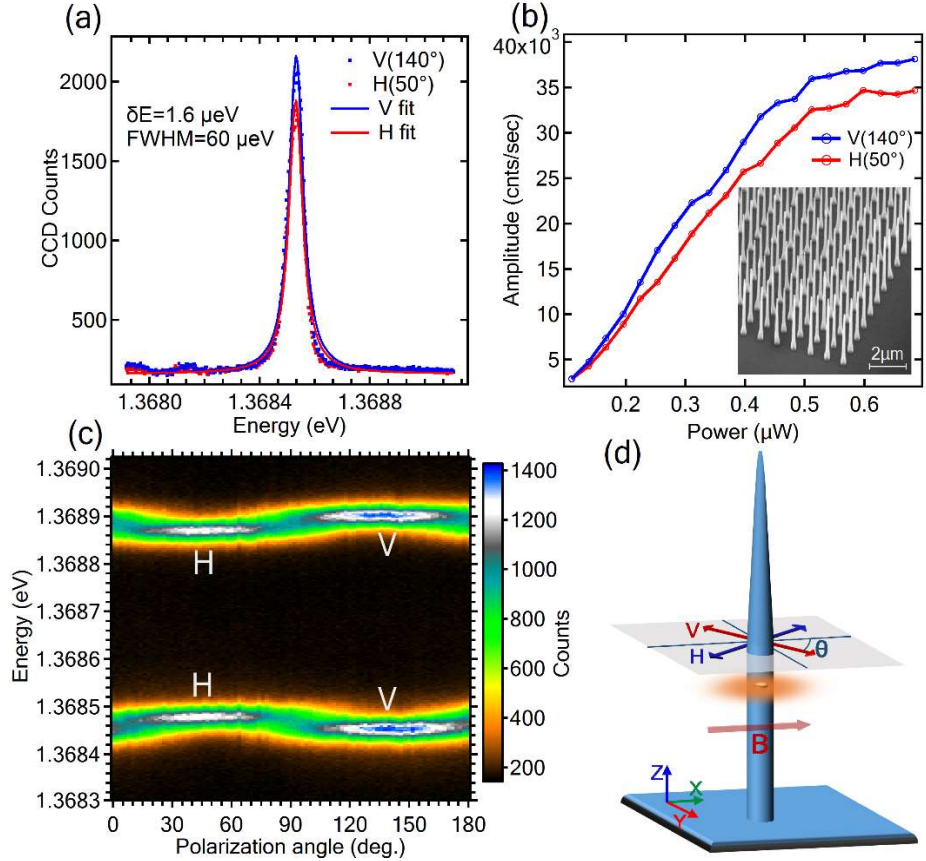

**Figure 1:** (a) QD photoluminescence spectrum, measured at $B = 0$ T, showing both the H- and V-polarized lines. (b) Dependence of the quantum dot photoluminescence intensity on the power of the above-band pump laser. Inset: SEM image of an array of site-controlled InP nanowires. (c) Polarization-resolved photoluminescence spectra from the quantum dot in a magnetic field of $B = 5$ T, showing four distinct spectral lines and their respective polarizations. (d) Main polarization orientations with respect to the orientation of the applied magnetic field. For this particular quantum dot, H is defined by $\theta_H=50°$ and V by $\theta_V=140°$.

We obtained further evidence that the QD was charged by recording photoluminescence spectra as a function of the applied magnetic field strength. Figure 2a shows the photoluminescence spectra of the quantum dot for varying magnetic fields in the range from 0 T to 5 T for both the H and V polarizations; the splitting of the emission into four lines is clearly evident. This is expected for a charged dot, for which there are two contributions to the transition energies that depend on the magnetic field: a linear dependence due to the Zeeman effect [27] (Figure 2b), and a quadratic dependence resulting from the diamagnetic shift [28],[29], (Figure 2c).

The four-fold splitting of the spectral lines, their linear dependence on the *B*-field (after the diamagnetic shift has been subtracted), their polarization properties, and their approximately equal brightness, indicate that the quantum dot has both the level structure and the selection rules of a charged quantum dot. Figure 2d shows the relevant energy level diagram for a charged quantum dot in a Voigt-configuration magnetic field[27], and the polarization selection rules for the optical transitions.

Figure 3 illustrates our spin-pumping experiments, and shows the main results. We performed a set of four different experiments, to demonstrate that we can perform spin pumping into either the $|\uparrow\rangle$ state or the $|\downarrow\rangle$ state, in each case using one of two (per spin state) available optical transitions. The inset of Figure 3a illustrates the quantum dot optical transitions we used for pumping and detection in one of the four experiments, and we use it here as an example case to describe the experiment in detail. The principle of the experiment is as follows [30],[31]. A fixed-wavelength above-band laser is used to randomize the state of the spin; it does this by incoherently exciting states at far higher energies than the QD trion levels, and through a series of decay processes, some of which are non-spin-preserving, the QD trion levels are randomly populated, and these levels in turn decay and randomly populate the QD ground state spin levels. The action of the spin randomization laser is depicted as violet upward wavy arrows. A tunable laser is used to resonantly excite one of the trion states via a vertical transition ($|\uparrow\rangle$-$|\uparrow\downarrow\Uparrow\rangle$). If the system is initially in the state $|\uparrow\rangle$, then this laser will cause the trion state $|\uparrow\downarrow\Uparrow\rangle$ to be populated. This trion state will then decay to either the $|\downarrow\rangle$ or $|\uparrow\rangle$ state with equal probability (grey downward wavy arrows in the inset of Figure 3a). If the decay is to the state $|\uparrow\rangle$, then the tunable laser will re-excite the trion state. If, on the other hand, the decay is to the $|\downarrow\rangle$ state, then the tunable laser will no longer be resonant with any transition and the system will be initialized in the $|\downarrow\rangle$ state, until the spin is randomized again. The emission from the $|\uparrow\downarrow\Uparrow\rangle$-$|\downarrow\rangle$ transition is spectrally filtered and sent to a single-photon counter, providing a measurement of the spin state[3].

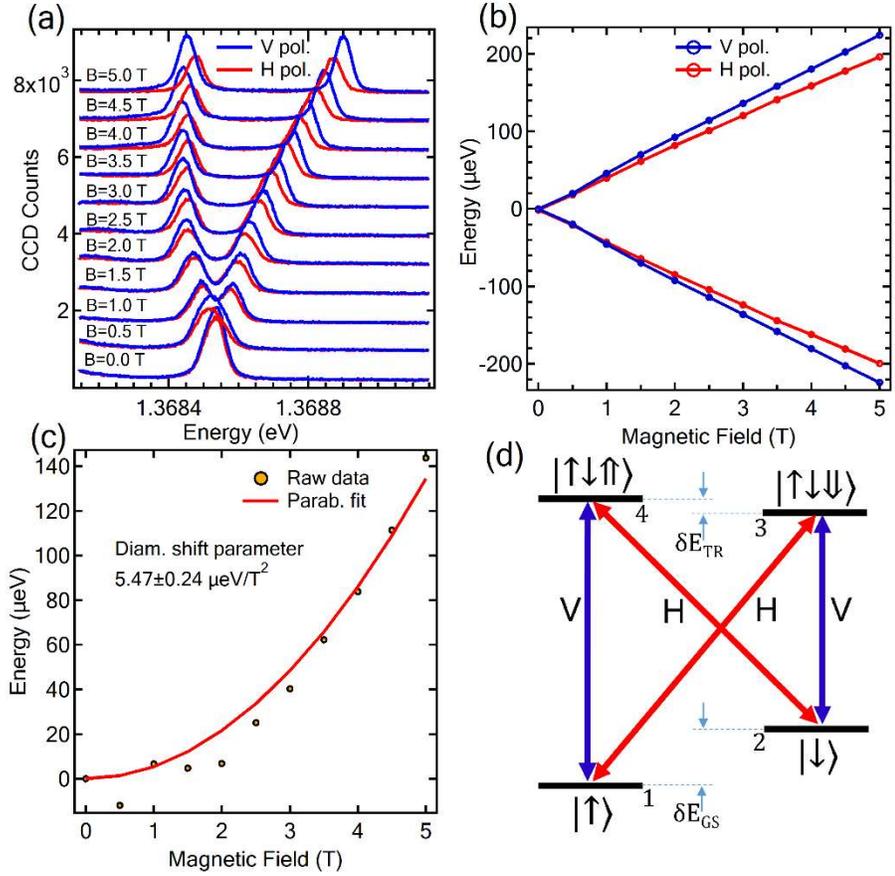

**Figure 2**. Voigt configuration magneto-photoluminescence spectra of the QD in a nanowire. (a) Raw spectra of photoluminescence at various magnetic fields B. (b) Energy shift of the optical transitions versus magnetic field, after subtraction of the diamagnetic shift. The linear dependence arises due to the Zeeman splitting of both the ground states and trion states. (c) The quadratic energy shift of the spectral lines due to the diamagnetic shift. (d) Energy level diagram of spin and trion states in a charged quantum dot, in a Voigt-configuration magnetic field (B perpendicular to the nanowire growth direction). All four optical transitions are allowed, and the polarization selection rules are indicated. $\delta E_{gs}$ is the ground state Zeeman splitting and $\delta E_{tr}$ the trion state Zeeman splitting.

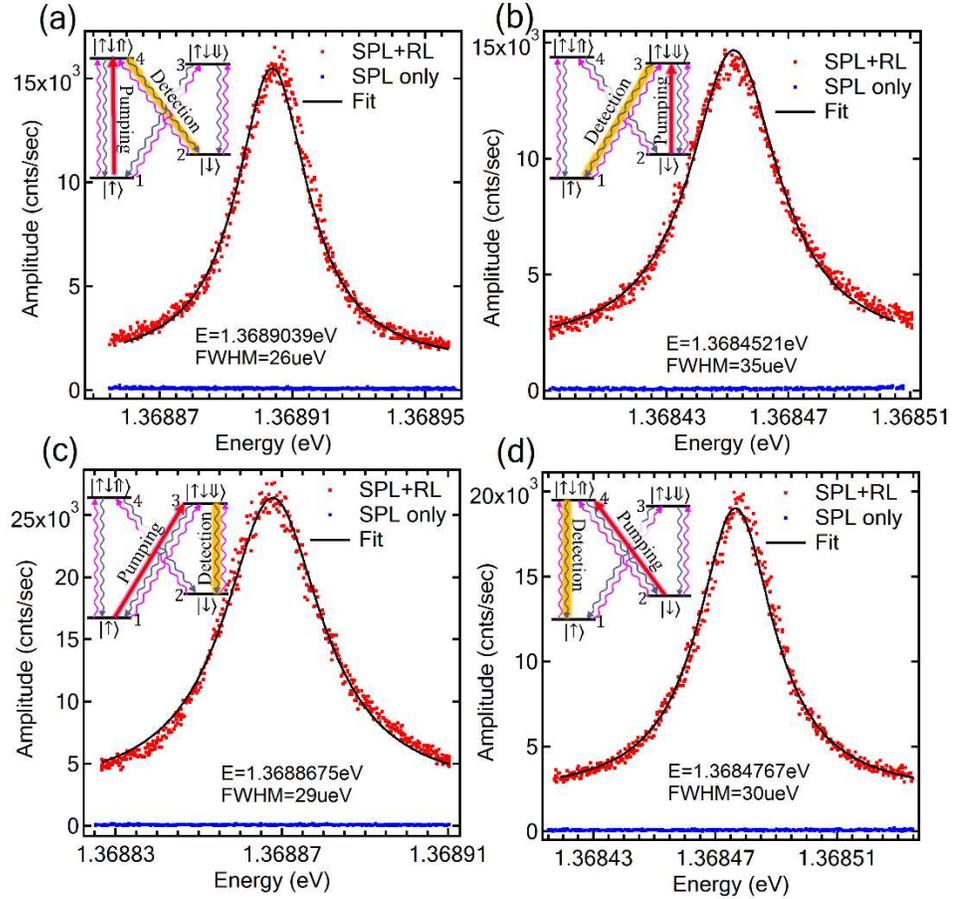

*Figure 3*. Spin-pumping and measurement experiment, for four different spin-pumping laser configurations. (a)-(d) Detected photons as a function of the spin-pumping-laser energy, for the specific experimental protocols illustrated in the respective insets. Red data points show the collected photon counts when both lasers were on, whereas the blue data points show the photon counts when only the spin-pumping laser was on. For all four spin pumping schemes the powers used were $P_{init}$ = 500 nW (spin-pumping laser) and $P_{rand}$ = 60 nW (above-band spin randomization laser). The insets are showing the relevant levels of the quantum dot, the applied laser fields, and photon collection. The spin randomization laser is shown as an upward violet wavy arrow, whereas spontaneous decay depicted as downward grey wavy arrows. A tunable CW laser ("Pumping"), is scanned across one of the optical resonances. Photons emitted by the yellow shaded transition ("Detection") are collected, and measured by a single-photon counter.

Figure 3a shows the collected photon counts as a function of the wavelength of the spin-pumping laser, as it was tuned over the $|\uparrow\rangle$-$|\uparrow\downarrow\Uparrow\rangle$ transition. Two traces are shown: one when only the spin-pumping laser was on (in blue), and one when both the randomization laser and the spin-pumping laser were on (in red). When both lasers were on, the data show a clear resonance, corresponding to spin-pumping-laser photons being absorbed by the $|\uparrow\rangle$-$|\uparrow\downarrow\Uparrow\rangle$ transition, and being emitted by the $|\uparrow\downarrow\Uparrow\rangle$-$|\downarrow\rangle$ transition. The reason that photons can be absorbed by the $|\uparrow\rangle$-$|\uparrow\downarrow\Uparrow\rangle$ transition is that the $|\uparrow\rangle$ state is continually being populated as a result of the randomization laser being on. However, when the randomization laser is turned off, the data shows no resonance as the spin-pumping laser passed over the $|\uparrow\rangle$-$|\uparrow\downarrow\Uparrow\rangle$ transition. This serves as strong evidence that the quantum dot spin has in this case been optically initialized in the $|\downarrow\rangle$ state.

In an analogous manner, Figure 3b shows how the quantum dot spin can be optically pumped into the $|\uparrow\rangle$ state via the other vertical optical transition ($|\downarrow\rangle$-$|\uparrow\downarrow\Downarrow\rangle$), and Figures 3c and 3d show how the quantum dot spin can be optically pumped using the two available diagonal optical transitions.

To demonstrate the robustness and repeatability of this spin qubit initialization technique in nanowire quantum dots, we performed the same type of optical pumping experiments on several nanowire quantum dots on the same sample, which yielded essentially identical results to those shown in Figure 3 (see Supplementary Data).

To further characterize the spin pumping process, we also studied the spin-pumping signal as a function of the applied spin-pumping-laser power. Figure 4 shows the dependence of the peak spin-pumping signal on the spin-pumping-laser power in the experiment in Figure 3a (where the spin is pumped into the $|\downarrow\rangle$ state using the $|\uparrow\rangle$-$|\uparrow\downarrow\Uparrow\rangle$ transition). In particular, the red data points in Figure 4a show that the peak of the spin-pumping signal initially increases rapidly with applied laser power, and saturates once the spin-pumping-laser power reaches approximately 350 nW. The blue data points are from the same experiment, except the randomization laser had been turned off. We note that even with a power well above the value that is sufficient to saturate the $|\uparrow\rangle$-$|\uparrow\downarrow\Uparrow\rangle$ transition (and hence cause maximal spin pumping), when the randomization laser is off, there is no increase in the counts as a function of power, which is consistent with high-fidelity optical pumping and excellent spectral-filtering-based rejection of the scattered spin-pumping-laser light. We also studied the effect of the applied laser power on the width of the transition resonance. Figure 4b shows the full-width-at-half-maximum linewidth of the resonance when both the randomization laser and the spin-pumping laser were on, as a function of the power of the spin-pumping laser. At lower powers ($P_{\text{spin-pump}} < 0.15$ μW), we observed a linewidth of less than 20 μeV, with substantial broadening as the power was increased to be above the saturation limit of the transition. The power dependence measurements in Figures 1b (above-band excitation photoluminescence power dependence) and 4a (spin-pumping laser power dependence) provide valuable information for quantitatively assessing the efficacy of the spin pumping process, since the saturation values can be used to infer the relative rates of spin pumping and spin randomization.

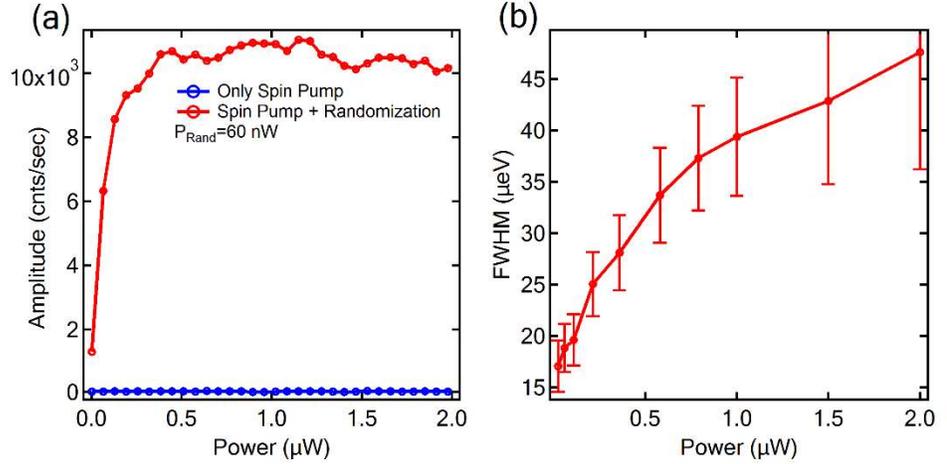

**Figure 4**. Spin-pumping peak amplitude and width as a function of the power of the spin-pumping laser, in the experiment described in Figure 3. The randomization laser power was kept constant at $P_{rand}$ = 60 nW. (a) Red: photon counts at the peak of the spin-pumping resonance (when both lasers were on), as a function of the spin-pumping laser power. Blue: photon counts measured when the randomization laser was off. (b) The full-width-at-half-maximum linewidth of the resonance shown in Figure 3a, as a function of the spin-pumping laser power.

We have used a rate-equation model (described in detail in the Supplementary Data) of the spin pumping experiment to analyze our experimental results; it shows that our data is consistent with optical spin pumping causing spin qubit initialization with a fidelity of 99% in less than 10 ns. This is similar to the reported performance of spin pumping in self-assembled InAs quantum dots[13],[16],[18]. We used our model to infer a lower bound on the spin lifetime of $T_1 > 3$ μs. These values suggest promise for the use of charges in nanowire quantum dots as spin qubits.

**Conclusion**. We have demonstrated that in the InAsP-QD/InP-nanowire system a charged quantum dot in a Voigt magnetic field does yield two optical Λ-systems that can be manipulated, and we have demonstrated optical spin pumping using independent experiments on both transitions in both Λ-systems. Furthermore, we were able to show that spin measurement as part of the optical pumping process is possible in the InAsP-QD/InP-nanowire system. These experiments were performed on site-controlled nanowires, making this the first demonstration of optical pumping of a site-controlled quantum dot spin.

**Corresponding Authors**. KGL (lagous@stanford.edu) or PLM (pmcmahon@stanford.edu).

**Funding Sources**. We acknowledge financial support from the Air Force Office of Scientific Research, the MURI Center for Multi-functional Light-Matter Interfaces based on Atoms and Solids, and from the Army Research Office (grant number W911NF1310309). This research was also supported by the Cabinet Office, Government of Japan, and the Japan Society for the Promotion of Science (JSPS) through the Funding Program for World-Leading Innovative R&D on Science and Technology (FIRST Program). KGL acknowledges support by the Swiss National Science Foundation. PLM was supported by a David Cheriton Stanford Graduate Fellowship.

**Acknowledgments**. We would like to thank Kristiaan De Greve for useful discussions.

**Abbreviations**. QD: quantum dot; InP: Indium Phosphide; InAsP: Indium Arsenide Phosphide, SCQD: site-controlled quantum dot; CW: continuous wave.

# Supplementary Data

# Initialization of a spin qubit in a site-controlled nanowire quantum dot

**Konstantinos G. Lagoudakis[1], Peter L. McMahon[1], Kevin A. Fischer[1], Shruti Puri[2], Kai Müller[1], Dan Dalacu[3], Philip J. Poole[3], Michael E. Reimer[4], Val Zwiller[5], Yoshihisa Yamamoto[1,6], Jelena Vučković[1]**

1 E.L. Ginzton Laboratory, Stanford University, Stanford, California 94305, USA

2 Département de Physique, Université de Sherbrooke, Sherbrooke, J1K 2R1 Québec, Canada

3 National Research Council of Canada, Ottawa, K1A 0R6, Canada

4 Institute for Quantum Computing and Department of Electrical and Computer Engineering, University of Waterloo, Waterloo, Ontario, N2L 3G1 Canada

5 Quantum Nano Photonics, KTH Royal Institute of Technology, SE-100 44 Stockholm, Sweden

6 National Institute of Informatics, Hitotsubashi 2-1-2, Chiyoda-ku, Tokyo 101-8403, Japan

**Quantum Dot Level Structure and Physical Parameters**

A charged quantum dot traps a conduction-band electron (or a valence-band hole), and so the ground state energy levels are the two spin states associated with that charge (which correspond to the spin of the trapped charge being aligned or anti-aligned with the applied magnetic field). The manifold of optically-excited states that are relevant in this paper are those which describe two different states of a single charged-exciton complex, which is known as a trion, and is a bound three-particle complex that consists of two electrons and one hole (or two holes, and one electron). Both the ground spin states and the trion states will in general exhibit a non-zero Zeeman splitting when a magnetic field is applied ($\delta E_{\text{gs}}$ and $\delta E_{\text{tr}}$), and this results in the creation of two $\Lambda$ systems ($|\uparrow\rangle$-$|\uparrow\downarrow\Uparrow\rangle$-$|\downarrow\rangle$ and $|\downarrow\rangle$-$|\uparrow\downarrow\Downarrow\rangle$-$|\uparrow\rangle$), where a single-line arrow denotes an electron spin eigenstate, and a double-line arrow denotes a hole spin eigenstate).

From Figure 2b, one can estimate the g-factors of the ground and excited state spins. We found the ground-state-spin g-factor to be $g_{gs}$=1.45±0.01, and the excited-state-spin g-factor to be $g_{es}$=0.1±0.01, which are similar to previously reported values[32]. By fitting a quadratic function of $B$ to the data in Figure 2c, we extracted a diamagnetic shift coefficient of 5.47±0.24 μeV/T$^2$, which is also similar to a previously reported value[32].

**Modeling**

Here we present our rate-equation model for extracting a conservative lower-bound on the lifetime of the ground spin states, $\tau_{|\downarrow\rangle,|\uparrow\rangle}$, i.e., the $T_1$ time of the spin qubit. This is a useful performance metric for a spin qubit on its own, but we are particularly interested in it here since the fidelity of our spin initialization is likely limited by spin lifetime. Due to the clear polarization selection rules of the quantum dot (shown in Figure 1c), we constrain our spin-pumping model to the populations of a single lambda system comprised of the two spin ground states and a single trion state ($|\uparrow\rangle$, $|\downarrow\rangle$, and $|\uparrow\downarrow\Uparrow\rangle$). We consider four rates in our system: spin-pumping ($R_s$), above-band spin-randomization ($R_{ab}$), spontaneous emission ($\gamma$), and ground-state spin flips ($\Gamma$). The spontaneous emission lifetime for excitons in quantum dots in nanowires has been measured in samples very similar to the one we used in our experiments, and has been found to be in the range 0.5 ns[33] to 2 ns[34],[35]. In our model, we assume a spontaneous emission lifetime of 1 ns. The spin-pumping and above-band pumping rates are known as fractions of their saturation powers, approximately 2 and 0.1 respectively. Since these rates are linear in their corresponding laser powers, these ratios give the spin-pumping rates relative to the spontaneous emission rate, i.e., $R_s = 2\gamma$ and $R_{ab} = \gamma/10$. Using this information, our goal is to infer from the spin-pumping experimental data in Figure 3 the spin-flip rate. This in turn allows us to specify a lower bound on the spin-state lifetime, and to estimate the fidelity of spin initialization.

We use the following rate-equation model, which disregards the quantum coherences between the three relevant states:

$$\frac{d}{dt}\begin{pmatrix} P_{|\uparrow\rangle}(t) \\ P_{|\downarrow\rangle}(t) \\ P_{|\uparrow\downarrow\Uparrow\rangle}(t) \end{pmatrix} = \begin{pmatrix} -\Gamma - R_{ab} & \Gamma & \gamma \\ \Gamma & -\Gamma - R_s - R_{ab} & \gamma + R_s \\ R_{ab} & R_s + R_{ab} & -2\gamma - R_s \end{pmatrix} \begin{pmatrix} P_{|\uparrow\rangle}(t) \\ P_{|\downarrow\rangle}(t) \\ P_{|\uparrow\downarrow\Uparrow\rangle}(t) \end{pmatrix}.$$

The state populations of the lambda system are given by $P_{|\uparrow\rangle}(t)$, $P_{|\downarrow\rangle}(t)$, and $P_{|\uparrow\downarrow\Uparrow\rangle}(t)$. Since our spin pumping experiments were performed with continuous pumping and non-time-resolved measurements, it is reasonable to assume that our measurement results were obtained when the system was in steady state. To solve for the steady state, we need only two of the above equations and the constraint that the total probability is conserved. In our model, for steady state, we obtain the following analytical results: when the spin-randomization laser is off, the trion population can be expressed as $P_{|\uparrow\downarrow\Uparrow\rangle}(t) \sim \Gamma/\gamma$; on the other hand, when the spin-randomization laser is on, $P_{|\uparrow\downarrow\Uparrow\rangle}(t) \sim R_{\text{ab}}/\gamma$. In the experimental results shown in Figure 3, our detected signal is $\propto P_{|\uparrow\downarrow\Uparrow\rangle}(t)$, so we can use the ratios between the peak signals when the above-band laser is on versus off to infer $\Gamma$. This ratio is limited by the dark counts of our detector (~ 50 counts/second). Using this method, we obtain a lower bound on the spin-state lifetime of $\tau_{|\downarrow\rangle,|\uparrow\rangle} \gtrsim 3$ μs. Under the assumption that the finite spin lifetime is the dominant limitation to obtaining perfect spin initialization fidelity using optical pumping, we can infer that a spin-initialization fidelity of 99% can be achieved in 10 ns.

**Experimental setup**

The experimental setup used for the spin pumping experiments is a standard confocal microscopy setup with high magnetic field capability as illustrated in Figure S1a. The sample is held at $T_{\text{exp}} \approx$ 8.5 K in a magnetic cryostat with maximum field $B_{\text{max}} = 5$ Tesla. The magnetic cryostat is a continuous helium flow superconducting Microstat MO manufactured by Oxford Instruments. The sample is optically accessed with a long working distance microscope objective (NA=0.5 and WD≈12 mm) and excited by two lasers: an above-band 780-nm diode (spin randomization laser) and a resonant single mode laser (spin pumping laser) by Coherent, model MBR, which allows for mode-hop free scans of up to 40GHz. The initialization laser frequency scans were performed by monitoring the single photon counts on an Excelitas SPCM at the output of the double monochromer while recording the exact wavelength of the laser using a high precision High Finesse WS-7 wavemeter by Angstrom. Spectral resolution is provided by a custom made 1.00m long monochromer coupled to an Acton 0.75-m monochromer. Both instruments feature a 1700ln/mm grating giving an overall resolution of about 10 μeV. The SPCM clicks are counted by an Arduino board using the frequency counting library by Martin Nawrath, nawrath@khm.de.

A more precise depiction of the geometry of the experiment is shown in Figure S1b. The magnetic field is applied in the Voigt configuration which is with the field lines perpendicular to the growth axis of the nanowire quantum dots. The magnetic field axis also serves as a reference for the polarization orientation and the detection angle.

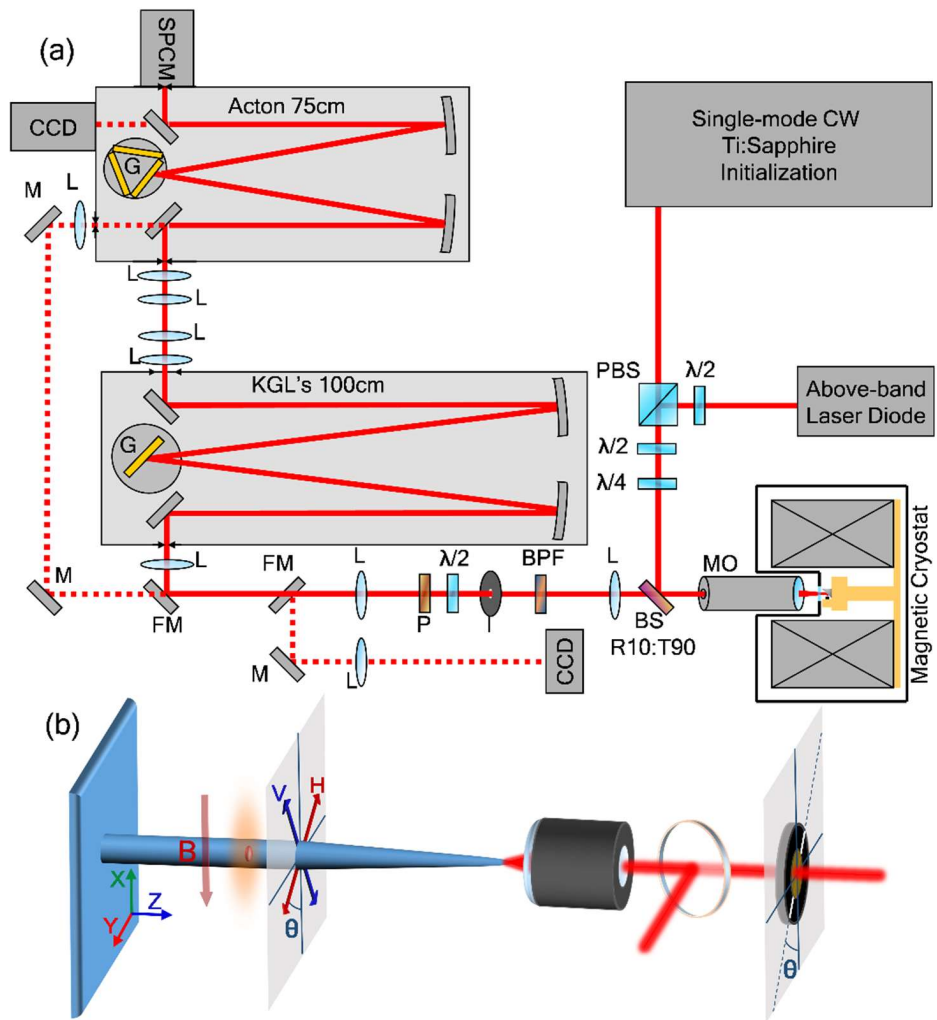

**Figure S1**: (a) Complete experimental setup for the spin pumping experiment. List of abbreviations: M: mirror, L: lens, MO: microscope objective, PBS: polarizing beam splitter, BS: beam splitter, BPF: band pass filter, I: iris, P: polarizer, FM: flip mirror, G: grating, λ/2: half

waveplate, λ/4: quarter waveplate, SPCM: single photon counting module, CCD: charge coupled device (camera). (b) Detailed illustration of detection scheme and polarization orientations with respect to magnetic field.

**Reproducibility**

To probe the robustness of the experimental demonstration of spin pumping, we performed the measurements on several nanowire QDs. For each QD we performed a spin pumping experiment using the schemes shown in Figure S2a-b. Charging of the dots was verified by the application of a magnetic field and the four-fold splitting of the QD photoluminescence spectra as shown in figure S2(c)-(f). Initially we pumped resonantly the vertical lowest energy transition $|\downarrow\rangle$-$|\uparrow\downarrow\Downarrow\rangle$ while collecting photons from the diagonal $|\uparrow\downarrow\Downarrow\rangle$-$|\uparrow\rangle$ (scheme (a)) shown in figure S2(g)-(j). We then pumped resonantly the diagonal $|\downarrow\rangle$-$|\uparrow\downarrow\Uparrow\rangle$ while collecting photons from the highest energy vertical $|\uparrow\downarrow\Uparrow\rangle$-$|\uparrow\rangle$ transition (scheme (b)) shown in figures S2(k)-(n). We performed these spin pumping experiments while both randomization and spin pumping laser were applied (red points) as well as when only the resonant spin pumping laser was on (blue points) showing excellent spin initialization characteristics. All four dots show identical behavior to the nanowire dot studied in detail in the main manuscript.

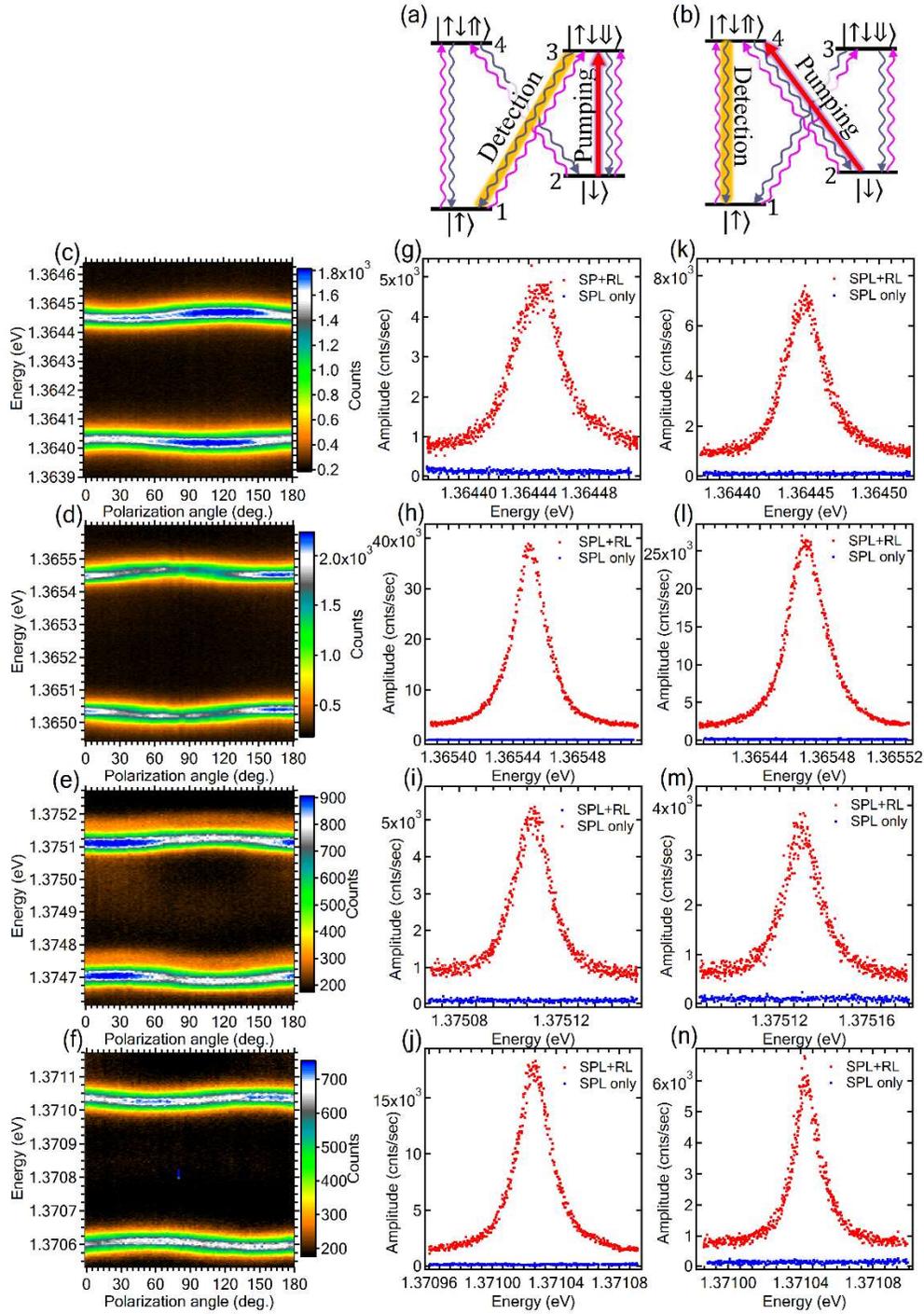

**Figure S2:** (a),(b) Two possible spin pumping schemes. (c)-(f) Polarization scans at B=5 T for four different nanowire QDs showing a fourfold splitting. (g)-(j) Spin pumping experiment according to scheme of panel (a). (k)-(n) Spin pumping experiment according to scheme of panel (b). All four QDs show excellent spin pumping characteristics with very good suppression when the randomization laser is not applied.

**Supplementary data Bibliography**